\documentstyle[twocolumn,aps,psfig]{revtex}
\begin{document}
\draft
\twocolumn[\hsize\textwidth\columnwidth\hsize\csname @twocolumnfalse\endcsname

\title{ The Phase Separation Scenario for Manganese Oxides}
\author{Adriana Moreo, Seiji Yunoki, and Elbio Dagotto}

\vskip 0.1cm

\address{National High Magnetic Field Lab and Department of Physics,
Florida State University, Tallahassee, FL 32306}

\date{\today}
\maketitle

\begin{abstract}

Recent computational
studies of models for manganese oxides 
have revealed a rich phase diagram, 
not anticipated in early calculations in this context performed in the
1950's and 60's. In particular, the transition
between the antiferromagnetic 
insulator state of the hole-undoped limit and the 
ferromagnetic metal at finite hole-density was found to occur
through a mixed-phase process. 
When extended Coulomb interactions are included, a microscopically charge
inhomogeneous state should be stabilized. 
These phase separation tendencies,
also present at low electronic densities, influence the properties of the
ferromagnetic region by increasing charge fluctuations. 
Experimental data reviewed here using
several techniques for manganites and other materials 
are consistent with this scenario. 
Similarities with
results previously discussed in the context of cuprates are clear from
this analysis, although the phase segregation tendencies in manganites
seem stronger.

\end{abstract}
\pacs{[To appear in Science]}

\vskip2pc]
\narrowtext

\vskip 0.8cm

{{\bf I. Introduction}}
\vskip 0.2cm

Hole-doped manganese oxides with a perovskite
structure have stimulated considerable
scientific and technological interest due to their exotic electronic
and magnetic properties~[1].
These manganites have a chemical composition
$\rm R_{1-x} A_x Mn O_3$, with R a rare-earth ion and A a
divalent ion such as Ca, Sr, Ba, or Pb.
They present an unusual magnetoresistance (MR) effect, whereby
magnetic fields induce
large changes in their resistivity $\rho$, a property that may find 
applications in sensor technologies such as that utilized in magnetic
storage devices. For example, 
in La-Ca-Mn-O thin films, the ratio
$\rm (\rho(0) - \rho(H))/\rho(H)$, with $\rm \rho(H)$ the
resistivity in a magnetic field H,
can be as large as $10^3$ at 77K (H=6T).
The term ``colossal'' magnetoresistance (CMR) has been coined
to describe this effect~[2].

The unusual properties of manganese oxides
challenge our current understanding of
transition-metal oxides, and define a  basic research
problem that involves an
interplay between the charge, spin, phononic, and orbital 
degrees of freedom.
Manganites have a  
rich phase diagram~[3] that includes a well-known
ferromagnetic (FM) phase that spans 
a robust range of electronic densities. 
The CMR effects have been observed particularly
at small hole densities
x but also at $\rm x$$\sim$$0.5$, which 
are the density 
limits of the FM-phase.
The strength of 
the MR effect increases as the
electronic bandwidth is decreased through chemical substitution~[1],
which also reduces the Curie critical temperature $\rm T_C$.
At hole concentrations $\rm x$$\sim$$0.5$, an
antiferromagnetic (AF)
charge-ordered (CO) insulating state, discussed 
by Goodenough~[4], is 
involved in the CMR effect,
which at these densities is extraordinarily large~[5].

In the undoped limit, the $\rm Mn^{3+}$ ions 
have four electrons in the 3d-shell, and they are surrounded
by oxygens $\rm O^{2-}$ forming an octahedron. This crystal
environment breaks the full rotational invariance,
causing the two $\rm e_g$- and three $\rm t_{2g}$-orbitals to split. 
The strong Hund
coupling ($\rm J_H$) in these systems favors the spin
alignment of the four
electrons in the active shell; on average  three electrons populate
the $\rm t_{2g}$-orbitals and one occupies the $\rm e_g$-states.
The $\rm t_{2g}$-electrons are mainly
localized, whereas the $\rm e_g$-electrons 
are mobile and use O p-orbitals
as a bridge between Mn ions. When the manganites are
doped with holes through chemical substitution,
$\rm Mn^{4+}$-ions with only three
$\rm t_{2g}$-electrons are formed. 
In addition, in the undoped limit
 the $\rm e_g$-degeneracy is split  due to Jahn-Teller (JT)
distortions; as a consequence, a one-orbital approximation has
been frequently used since the earliest theoretical studies~[6]. 
For these reasons, typical
electronic models
for the manganites include at least a kinetic energy contribution
for the
$\rm e_g$-electrons, regulated by a hopping amplitude $\rm t$, and 
a strong $\rm J_H$ coupling contribution
between the $\rm e_g$- and $\rm t_{2g}$-spins.
The localized spin is large enough (3/2) 
to be approximated by a classical spin,
which simplifies the calculations. 
Here this model will be simply
 referred to as the ``one-orbital'' model, although other names
are sometimes used, such as FM Kondo model.

This formalism leads to a natural explanation for the
FM-phase of the manganites, because
carriers energetically prefer to polarize the spins in their vicinity.
When an $\rm e_g$-electron jumps between nearest-neighbor ions,
it does not pay an energy $\rm J_H$ if
all of the spins involved are parallel. 
The hole-spin scattering is minimized in this
process, and the kinetic energy of the mobile carriers is optimized.
This mechanism is usually referred to as double-exchange (DE)~[6,7].
As the carrier density grows, the FM distortions
around the holes start overlapping and 
the ground state becomes fully ferromagnetic.

 Currently there is not much controversy about the qualitative
 validity of DE to stabilize a FM-state.
However, several experimental results suggest
that more complex ideas are needed to explain the main 
properties of manganese oxides. 
 For instance, above $\rm T_C$ and for a wide range of densities, several
 manganites exhibit
 insulating behavior of unclear origin
that contributes to the large MR results.
 The low-temperature (T) phase diagram of these materials has a complex 
structure~[3], not predicted by DE,
that  includes insulating AF- and CO-phases, orbital
ordering, FM-insulating regimes, 
and, as discussed extensively below,
tendencies toward the formation of charge inhomogeneities, even within
the FM-phase.
To address the strong MR effects and
 the overall phase
diagram of manganites, the DE framework must be
supplemented with more refined ideas.

\vskip 0.5cm
{\bf II. Phase Separation in the One-Orbital Model}
\vskip 0.2cm

 Motivated by new  experimental research on manganese oxides,
there has been considerable theoretical
work in the analysis of models for these materials.
Several many-body techniques for modeling strongly correlated 
electron systems were developed and improved during recent 
efforts to understand high-temperature superconductors, and thus it
is natural to apply some of these methods to manganite models.
 Of particular relevance here are the computational
 techniques that allow for an unbiased analysis of correlated models
 on finite clusters~[8].
The first comprehensive computational analysis of
the one-orbital model was 
presented by Yunoki et al.~[9] using classical spins for the 
 $\rm t_{2g}$-electrons and the Monte Carlo (MC) technique. 
Several unexpected results were found in this study. In particular,
when calculating the density of $\rm e_g$-electrons 
$\rm \langle$$\rm n$$\rm \rangle$ $\rm =(1-x)$ 
as the chemical potential $\mu$ was varied, it was surprising to observe that
some densities could not be stabilized;
in other words, 
$\rm \langle$$\rm n$$\rm \rangle$ was found to change
discontinuously at
special values of $\mu$. These densities are referred to as ``unstable.''
Alternative calculations in the
canonical ensemble~[9,10] where the density is
fixed to arbitrary values, rather
 than being
regulated by $\mu$, showed that at
unstable densities,
the resulting ground state is $not$ homogeneous, but it is separated
into two regions with differing densities.
The two phases involved correspond to those that bound the unstable
range of densities~[9-11].
This phenomenon, which has been given the name of
``phase separation'' (PS), appears in many contexts, such as the familiar
liquid-vapor coexistence in the phase
diagram of water, and it is associated
with the violation of the stability condition $\rm \kappa^{-1} = $$
\rm \langle n \rangle^2$$\rm \partial^2 E / \partial \langle n \rangle^2 > 0$, 
with $\rm E$ the energy of the system per unit volume, 
and $\kappa$ the compressibility.

In the realistic limit $\rm J_H/t$$\rm \gg$$\rm 1$,
phase separation occurs between hole-undoped 
$\rm \langle$$\rm n$$\rm \rangle$$=1$ 
and hole-rich 
$\rm \langle$$\rm n$$\rm \rangle$$<$$1$ 
phases~[9-11]. Although the 
$\rm e_g$- and $\rm t_{2g}$-spins of the same ion
tend to be parallel at large $\rm J_H$, their relative orientation at 
one lattice-spacing depends on
the density. At 
$\rm \langle$$\rm n$$\rm \rangle$$=$$1$,
an AF arrangement results because
the Pauli principle precludes movement of the
electrons if all spins are aligned. However, at stable 
$\rm \langle$$\rm n$$\rm \rangle$$<$$1$ 
densities, DE forces the spins to
be parallel, as computer studies have indicated~[9-11]. 
Yunoki and Moreo~[11] have shown that if an additional small
Heisenberg coupling among the localized spins is introduced,
PS occurs also at small $\rm \langle$$\rm n$$\rm \rangle$,
this time
involving FM- 
($\rm \langle$$\rm n$$\rm \rangle$$>$$0$)
and electron-undoped AF-states. 
Phase segregation 
near the hole-undoped and fully-doped limits
implies that a
spin-canted state~[6] for the one-orbital model
is {\it not} stable.
Other authors arrived to similar conclusions
after observing phase segregation tendencies 
using several analytical techniques~[12-14]. If a spin-canted state is
unequivocally found in experiments, mechanisms other than DeGennes'~[6] 
may be needed to explain it. Note also that
a canted state is difficult to distinguish experimentally
from a mixed AF-FM state.

It is interesting that PS behavior is not unique to manganese
oxides. Indeed, 
the existence of PS in AF
rare-earth 
compounds has been addressed by
Nagaev for many years~[15]. 
In these materials, there is a small
density of electrons interacting with localized spins. Actually,
$\rm Eu_{1-x} Gd_x Se$ 
has a very large MR effect similar to that observed in manganites~[16,17].
Calculations in
this context were performed using the one-orbital model mainly
in the limit where the localized-conduction
spin-spin coupling is smaller than the bandwidth (equivalent to
$\rm J_H$$\ll$$\rm t$) and at small
$\rm \langle$$\rm n$$\rm \rangle$.
Nevertheless, some of these results have been discussed also in the context
of manganites~[18]. Note that in the
 recently established phase diagram  of the one-orbital model,
PS occurs
at {\it both} high- and low-electronic density~[11].
Because the $\langle {\rm n} \rangle$$\ll$$1$ 
limit corresponds to the  dilute AF semiconductors mentioned above,
the computational studies confirm that these materials
should also exhibit PS tendencies.
A broad distribution of
FM cluster sizes should be expected, with a concomitant distribution of
electrons trapped in those clusters~[19].
Gavilano et al.~[20] have recently reported a
two-phase mixed regime in these materials that may be related to 
intrinsic PS tendencies.
Analogous results were also observed in other diluted magnetic
semiconductors~[21].
The analysis of experimental
data in this context should certainly 
allow for the possibility of large-scale
inhomogeneous states. 

\vskip 0.5cm
{\bf III. Phase Separation in the Two-orbitals Model}
\vskip 0.2cm

Most of the theoretical studies for manganites have been carried out using
the one-orbital model, which certainly provides a useful playground for
the test of qualitative ideas. However, quantitative calculations
must necessarily include two active $\rm e_g$-orbitals per Mn-ion 
to reproduce
the orbital-ordering effects known to occur in
these materials~[22]. In addition, 
it has been
argued
that dynamical JT effects cannot be neglected~[23],
and the electron-JT-phonon 
coupling $\lambda$ should be important for the manganites.

\begin{figure}[htbp]
\vspace{-0.8cm}
\centerline{\psfig{figure=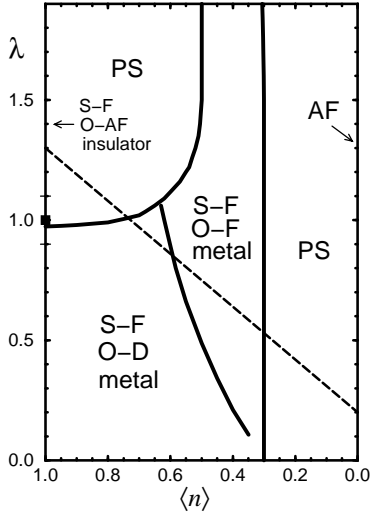,width=6.50cm,angle=-0}}
\vspace{0.0cm}
\caption{
Phase diagram of the two-orbital model for manganites in
one-dimension and $\rm T \sim 0$ including Jahn-Teller phonons,
 obtained with Monte Carlo techniques~[24].
S-F labels a spin-ferromagnetic configuration.
O-F, O-AF, and O-D denote a state where the orbital degrees of freedom
are ordered uniformly, staggered or they are disordered, respectively;
PS indicates a phase separated state, and AF is an antiferromagnetic state.
The Hund-coupling is $\rm J_H =
8$, the Heisenberg coupling between localized classical spins
$\rm J'=0.05$, both in units of the hopping
among the same orbitals. 
The meaning of the dashed line is explained in the text.
}
\end{figure}

Although computational 
studies accounting for JT-phonons are at their
early stages, some illustrative results are already available. 
Yunoki et al.~[24] recently reported the low-temperature phase
diagram of a two-orbital model 
using the Monte Carlo method, and analyzed the
results in a manner similar to the one-orbital case.
The results are reproduced in Fig.1 for a one-dimensional (1D) system at large 
Hund coupling.
The phase diagram is rich and includes a
variety of phases such as metallic and insulating regimes
with orbital
order. The latter can be uniform, with the same combination of
orbitals at every site, or staggered, with two
combinations alternating between the even- and odd-sites of the 
lattice at $\langle n \rangle = 1$. Recently, our group observed that
the density of states exhibits
$pseudogap$ behavior caused by the PS tendencies, both in the one- and
two-orbital cases, in agreement with photoemission experiments for
layered manganites~[25]. 
Of special
importance for the discussion here are the regions of unstable
densities. Phase separation appears at small
$\rm e_g$-densities
between an electron-undoped AF-state and a metallic 
uniform-orbital-ordered 
FM-state. The latter phase itself coexists at larger densities and
intermediate values of $\lambda$ with an
insulating 
($\rm \langle$$\rm n$$\rm \rangle$$=$$1$) 
staggered-orbital-ordered FM-state, in 
an orbital-induced PS process~[24].
The overall results are
qualitatively similar to those obtained with other model parameters, and 
in studies of 2D and 3D systems.
Overall, PS
tendencies are 
strong both in the one- and two-orbital models, and over a 
wide range of couplings. 
Similar 
tendencies have been recently
observed including large
on-site Hubbard interactions~[26], which is
reasonable because at intermediate and large electron-phonon coupling
a negligible probability
of on-site double-occupancy was found~[24].

The macroscopic
separation of two phases with 
different densities, and thus different charges, should actually
be prevented by long-range
Coulombic interactions, 
which were not incorporated into the one- and two-orbital 
models discussed thus far. 
Even including
screening and polarization effects,
 a complete separation leads to a huge energy
penalty. This finding immediately suggests that the two large regions
involved in the process
will break into smaller pieces to spread the charge more uniformly.
These  pieces are hereafter referred to as polarons
if they consist of just one-carrier in a local environment that has been
distorted by its presence. This distortion can involve 
nearby spins (magnetic polaron), nearby ions
(lattice polaron), or both, in which case this object will be simply
referred to as a ``polaron.'' However,
the terms ``clusters'' or ``droplets'' are
reserved for extended versions of the polarons,
characteristics of a PS regime,
 containing several
carriers inside a common large magnetic distortion or lattice distortion
or both.

The present discussion 
suggests that in the regime of unstable densities
the inclusion of
extended Coulomb interactions will lead to a stable
state, with clusters of one phase embedded in the other [see also (15)].
It is expected that the competition between 
the attractive DE tendencies among carriers and the Coulomb
forces will determine the size
and shape of the resulting clusters.
Either sizable droplets or polarons 
 may arise as the most likely
configuration~[27].
The stable state resulting from the inclusion of extended Coulomb interactions
on an otherwise PS unstable regime will be referred to as a
charge-inhomogeneous (CI) state. 
However, the ideas presented here will still be described as the ``PS
scenario,'' with the understanding that only microscopic phase segregation
is the resulting net effect of the DE-Coulomb competition. Related ideas
have been previously discussed in the context of the cuprates~[28], with
attractive interactions generated by antiferromagnetism or phonons.
An exception to the existence of only purely
 microscopic effects occurs if the competing phases have 
approximately the same density, as observed experimentally at 
$\rm x = 0.5$ (discussed below). In this case, large-scale PS can be
expected.
Note also that the CI-state is certainly different from 
the metastable states that arise in a standard first-order
transition.
Figure 2 contains a cartoon-like 
version of possible charge arrangements in the CI-state, 
which are expected to fluctuate in shape, especially at high temperature
where the clustering is dynamic.
Unfortunately, actual calculations supporting a
particular distribution are still lacking. 
Nevertheless, the presently available results are sufficient to establish
dominant trends and to allow
a qualitative comparison between theory and experiment,
as shown below.

Phase separation 
in manganese oxides has clear
similarities with the previously discussed charge inhomogeneities 
observed in copper and nickel oxides~[28]. 
Actually, studies of 1D
 generalizations of the t-J model by Riera et al.~[29] showed that
as the localized spin magnitude S grows, 
the phase diagram is increasingly dominated
by either FM or PS tendencies. The importance of PS arises from
the dominance of the Heisenberg interactions over the kinetic
energies as S increases, which causes holes to be expelled from the 
AF-regions because they damage the spin environment. 
The tendency toward phase segregation decreases across the
transition-metal-row, from a strong tendency in Mn, to a weak tendency 
in Cu~[29].
The stripes observed in
cuprates~[28] could certainly appear in manganites as well through the
competition of the DE 
attraction and Coulomb repulsion among clusters.

\begin{figure}[htbp]
\vspace{-1.5cm}
\centerline{\psfig{figure=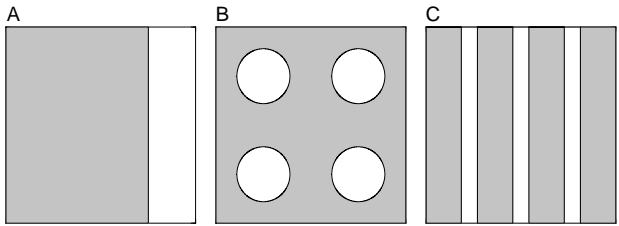,width=8.5cm,angle=-90}}
\vspace{-1.0cm}
\caption{
{Qualitative representation of (A}) a macroscopic phase separated 
state, as well as possible charge-inhomogeneous states stabilized 
by extended Coulomb interactions, such as (B) spherical droplets 
or (C) stripes. The diameter of the droplets and stripes are 
expected to be microscopic, but realistic calculations are 
still lacking.
}
\end{figure}

\vskip 0.5cm
{\bf IV. Influence of Phase Separation on the Ferromagnetic Phase}
\vskip 0.2cm

A critical aspect of the scenario discussed here
is the influence that
the low-temperature PS regime exerts 
on the behavior of electrons at higher temperatures,
and especially on the ordered phases which  
neighbor PS regimes.
As an illustration, consider the lines of
constant 
$\kappa$$\rm \langle n \rangle^2 =$$\rm d\langle n \rangle/d\mu$
of the 1D one-orbital model at large $\rm J_H$ (Fig.3A).
Because PS occurs through the 
divergence of $\kappa$, naturally
this quantity is the largest at those densities where
PS is observed (see above). A large $\kappa$ implies that
strong charge fluctuations occur, because $\kappa$$\rm ~\propto
( \langle {N}^2 \rangle - \langle { N} \rangle^2 )$, 
with $\rm { N}$ the total
number of particles.
A characteristic crossover temperature 
for ferromagnetism $\rm T^*_C$, occurs where
the zero-momentum spin structure factor 
starts growing very rapidly as the temperature is reduced. 
$\rm T^*_C$ is expected to become truly  critical
in higher dimensional systems, where also a finite critical temperature for
PS is expected to exist. Figure 3A shows  
that the PS tendencies influence the neighboring FM-state
because the compressibilities close to the PS regime,
located at $\rm 0.8~$$\leq$$\rm \langle n \rangle$$\leq 1.0 $,
are much larger than those
at, e.g., $\rm \langle n \rangle$$=0.5$.
This result implies that 
even within the FM-phase, which is uniform when time-averaged, 
there is a
dynamical tendency toward cluster formation because $\kappa$ is large.
The same situation occurs for $\rm T$$>$$\rm T^*_{C}$ and at low 
hole-densities.

\begin{figure}[htbp]
\vspace{-0.2cm}
\centerline{\psfig{figure=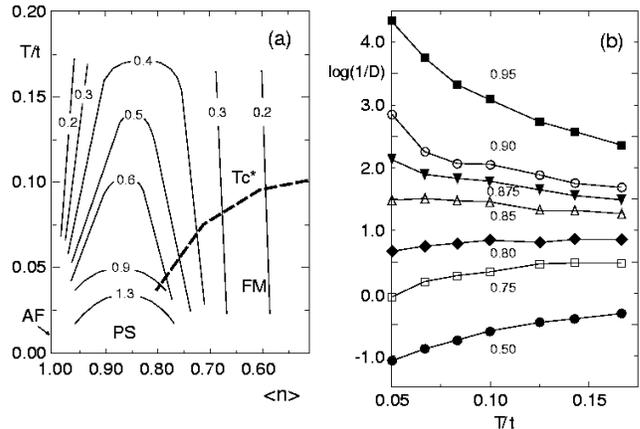,width=8.9cm,angle=-90}}
\vspace{0.0cm}
\caption{
(a) Lines of constant $\rm \kappa \langle n \rangle^2$$\rm
= d\langle n \rangle/d\mu$ that correspond to the
one-orbital model with $\rm J_H = \infty$ in the plane 
temperature-$\langle n \rangle$ ($\rm e_g$-density).
These plots were obtained
from Monte Carlo simulations using a 30-sites chain with
periodic boundary conditions.
The coupling among the localized $\rm t_{2g}$-spins is $\rm J'=0.05t$,
and $\rm t$ is the hopping amplitude. $\rm T^{*}_C$ is a characteristic
temperature where the zero-momentum Fourier-transform of the
 spin-spin correlations
starts growing very rapidly as the temperature is reduced~[9-11]; (b) A
plot of $\rm
log(1/D)$ versus $\rm T/t$ for the same model and cluster as in (A). $\rm D$ is the
Drude weight obtained from the optical conductivity~[11].
The densities are indicated.
}
\end{figure}

This effect should influence 
transport properties, including resistivity.
Although it is difficult
to evaluate $\rho$ with finite-cluster 
techniques, crude estimations can be made using
the inverse of the zero-frequency Drude weight found from
the optical conductivity~[11]. 
As an example, results are shown in Fig.3B for the one-orbital
model. This  estimation of $\rho$ produces the
qualitatively expected results, namely, it behaves as an insulator at
small x and rapidly decreases as x increases, turning smoothly
into a metal. Studies 
by our group using more sophisticated techniques connecting the cluster
with ideal metals have recently been found to produce qualitatively
similar data.
The results compare well with
experiments for Sr-doped compounds~[30]. 
Starting from a regime with dynamical cluster formation
above $\rm T_C$, the metallic state can be obtained 
if the clusters grow in size as T is reduced, eventually reaching
the limit where percolation is possible. At this temperature, the
carriers move over long distances and the metallic state is reached.
The same mechanism arises in polaronic theories~[31] [see also (32)].

\vskip 0.5cm
{\bf V. Comparing Theory with Experiments: the Phase Diagrams}
\vskip 0.2cm

 The computational results are consistent
with several experiments
on a variety of manganese oxides. 
Consider, for example, $\rm La_{1-x} Sr_x Mn O_3$. 
The experimentally observed sharp increase in
$\rho$ as $\rm x$ decreases towards the undoped limit,
both above and below
$\rm T_C$~[30], is difficult to explain if the only effect of the
correlations were to induce a reduced effective electronic hopping
$\rm t_{eff} = t \langle cos(\theta/2) \rangle$, where $\theta$ is the angle
between nearest-neighbor sites~[6].
In addition, the insulating properties of the
intermediate region $\rm 0.0 <  x < 0.16$~[30]
do not fit into the simpler versions of the DE ideas.
This regime is important because the CMR effect is maximized at
the lowest $\rm T_C$, that is, at the boundary between the
metallic and insulating regions.
Note also that recent experiments~[33,34] 
for hole-densities slightly above x=0.5 showed that 
the ground state of the $\rm (La,Sr)$-based manganese oxide
is an A-type AF-metal with
uniform $\rm d_{x^2 - y^2}$ orbital-order (Fig.4A). This 
phase, as well as the orbital-ordered x=0 A-type AF-insulator, 
does not appear in the one-orbital model~[9].
For these reasons, $\rm La_{1-x} Sr_x Mn O_3$ does  not seem
a typical DE material when considered including
 all available  densities, although at 
$\rm x \sim 0.3$ Furukawa~[35] showed that it has DE characteristics.

However, the experimental results mentioned above
can be more naturally accounted for once PS
 tendencies and the coupling with
JT-phonons are considered (narrower band materials are more complicated 
because of their x=0.5 CO-state, and they will be discussed below).
Actually, it has already been argued that
$\rho$ should rapidly grow as $\rm x$ 
decreases due to the strong charge fluctuations
at small $\rm x$ caused by the nearby phase 
segregation regime (Fig.3B).
In this context,
the insulating state above $\rm T_C$ of the lightly hole-doped
$\rm (La,Sr)$-compound can be
rationalized 
as formed by clusters of one phase (FM or AF) embedded into
the other.
Even the experimentally observed 
A-type AF-metallic
$\rm d_{x^2 - y^2}$-ordered phase  at $\rm x \sim 0.5$ (Fig.4A)
can be related to the phase with similar  characteristics 
near x=0.5 found in the theoretical calculations (Fig.1).
Although simulations with 
the realistic hopping amplitudes needed to stabilize an A-type AF-state
in a 3D environment have not  been performed yet, 
at least the 1D and 2D FM tendencies,
as well as the stabilization of a uniform orbital-ordering 
(Fig.1), are clear.
If phenomenologically one assumes
that $\rm \lambda/t$ decreases with hole doping,
the dashed-line in Fig.1 runs through the proper series of experimentally
observed phases~[33], namely, an
insulating staggered orbital-ordered state at x=0, a charge-segregated
regime at small x, a metallic orbital-disordered FM-phase at a higher
density, and finally
the $\rm x \sim 0.5$ orbitally ordered FM-state
compatible with A-type AF-order in dimensions $\rm D < 3$~[36]. 
If it were possible to complete the phase diagram of
$\rm La_{1-x} Sr_x Mn O_3$ by synthesizing $\rm x$$>$$0.6$ samples,
the calculations predict a new mixed-phase, involving
the A-type AF-metal with $\rm x<1$ and a G-type AF-insulator with $\rm x=1$,
where large MR effects could
potentially occur.

\begin{figure}[htbp]
\vspace{-0.0cm}
\centerline{\psfig{figure=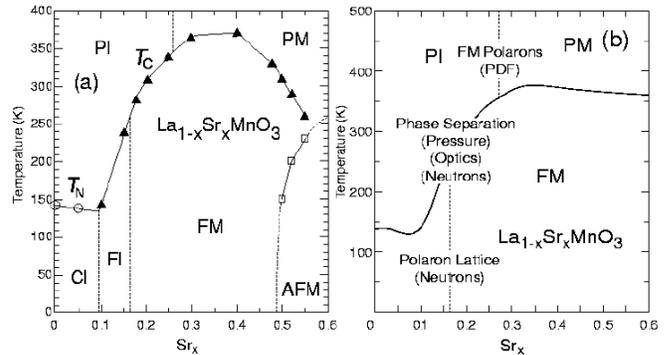,width=9.1cm,angle=90}}
\vspace{-0.3cm}
\caption{
(a) Phase diagram of $\rm La_{1-x} Sr_x Mn O_3$, courtesy of
Prof. Y. Tokura and Dr. Y. Tomioka, prepared with data taken from
Ref.~[30] and from Ref.~[68].
The ``AFM'' phase at large x is an A-type AF-metal
with uniform orbital-order. In this material there is no 
charge-ordered AF-state near x=0.5. PM, PI, FM, FI and CI denote
paramagnetic metal, paramagnetic insulator, ferromagnetic metal,
ferromagnetic insulator, and spin-canted insulator states, respectively.
$\rm T_C$ is the Curie temperature and $\rm T_N$ the N\'eel temperature;
(b) Schematic version of the phase diagram of $\rm La_{1-x} Sr_x Mn O_3$
to illustrate the experimental observation
of the tendencies  for
charge inhomogeneities in this material.
The language used  
to describe these tendencies, and the experimental techniques 
employed, are indicated
at the proper densities and temperatures. 
Additional details and references
can be found in the text. 
}
\end{figure}

\vskip 0.5cm
{\bf VI. Experimental Evidence of Charge Inhomogeneities}
\vskip 0.2cm

 Independently of 
the development of the theoretical ideas on PS, 
a large body of experimental evidence
 has accumulated that suggests the
 existence of charge inhomogeneities in manganese oxides either in macroscopic
form or, more often,
through the presence of small
 clusters of one phase embedded into another. The results have 
 been obtained on several materials,
at a variety of
 temperatures and densities, and using
 a large array of microscopic and macroscopic
 experimental techniques. 
These
 studies have individually concentrated on particular parameter regions, 
 and the results have rarely been discussed in comparison to similar
results obtained in other phase regimes. 
However, once all of these experimental data are combined, it appears
that the manganite metallic FM-phase
is surrounded both in temperature and density
 by charge inhomogeneous regions involving FM
clusters coexisting with another phase, which in some cases is AF. 
It would be unnatural to search for special
justifications for each one of these experimental results. The most
economical hypothesis is to explain the data 
as arising through a single effect, such as tendencies 
 to PS that compete strongly with ferromagnetism
both at large and small x, as well as above $\rm T_C$.
The experimental details are the following:

\vskip 0.3cm

{\bf VI.1 Sr-doped Manganese Oxides:}
Part of the phase diagram of 
$\rm La_{1-x} Sr_x Mn O_3$ is schematically shown in Fig.4B, including
at their proper location in T and $\rm \langle n \rangle$
some of the descriptions of charge inhomogeneity found in the literature,
along with the experimental techniques that have
reported such inhomogeneity.
They include results by Egami et al.~[31], 
where evidence for an inhomogeneous FM-state and
 small polarons both at high-
 and low-T 
was reported using pair-density
functional (PDF) techniques. In addition,
a recent analysis of the optical conductivity 
of $\rm La_{7/8} Sr_{1/8} Mn O_3$ by Jung et al.~[37] observed
PS features in the data.
Magnetic, transport, and
neutron scattering experiments by Endoh et al.~[26] on
$\rm La_{0.88} Sr_{0.12} Mn O_3$ 
revealed PS tendencies between two 
FM-phases, one metallic and the other insulating.
Other authors have also reported inhomogeneities in
Sr-doped compounds~[38].

\vskip 0.3cm

{\bf VI.2 Ca-doped Manganese Oxides:}
 In Fig.5, the phase diagram of 
 $\rm La_{1-x} Ca_x Mn O_3$ is sketched. Although previous reports~[3]
contain more details, for the present discussion just the
dominant qualitative aspects are needed. 
The list is not exhaustive but it is sufficient to illustrate
the notorious presence of charge inhomogeneities in this compound near
the FM-phase, and even inside it.
Overall, the analysis of available data for
  Ca-doped $\rm La Mn O_3$ leads to conclusions 
 similar to those presented for their Sr-doped counterpart.

The details are the following (postponing the
analysis for $\rm x$$\sim$$0.5$, which requires
special discussion).
Consider first the small-angle
 neutron scattering (SANS) results at x=0.05 and 0.08  and low-T
by M. Hennion et al.~[39]. They
 revealed the existence of a liquid-like distribution of
 FM droplets with a density
1/60 that  of holes.
In a similar regime of parameters, 
nuclear magnetic resonance (NMR) experiments by 
Allodi et al.~[40] reported the
coexistence of FM and AF features and the 
absence of spin-canting~[41]. 
SANS results 
at x=1/3 and $\rm T > T_C$ by Lynn et al.~[42]
and De Teresa et al.~[43] observed a short (weakly 
T dependent)  
FM correlation length,
attributed in~[43] to 
magnetic clusters $\rm 10$ to $\rm 20 \AA$
in diameter. Other experimental results, not reviewed here, agree
with this conclusion. 
Even within the metallic FM-phase ($\rm T < T_C$), indications of
 charge inhomogeneities have been reported. 
Transport measurements by Jaime et al.~[44] were
analyzed using a two-fluid picture
involving polarons and free electrons.
$\mu$-Spin relaxation and
resistivity data by Heffner et al.~[45] were interpreted as produced
by a multidomain sample.
X-ray absorption results by Booth et al.~[46] provided
evidence of coexisting localized and delocalized holes  below
$\rm T_C$~[47]. 
Using Raman and optical spectroscopies, S. Yoon et al.~[17]
found localized states in
the low-T metallic FM-phase of several manganese oxides.
Neutron scattering experiments~[42] reported 
an anomalous diffusive component in the data below
$\rm T_C$, which could be explained by a two-phase state.
Fernandez-Baca et al.~[48] have shown that this diffusive component
is enhanced as the $\rm T_C$ of the considered manganite 
decreases.
Actually, the low-energy component of the
two-branch spin-wave spectrum observed at small x~[49] 
has similarities with the diffusive peak at
x=1/3~[42].

\begin{figure}[htbp]
\vspace{-0.2cm}
\centerline{\psfig{figure=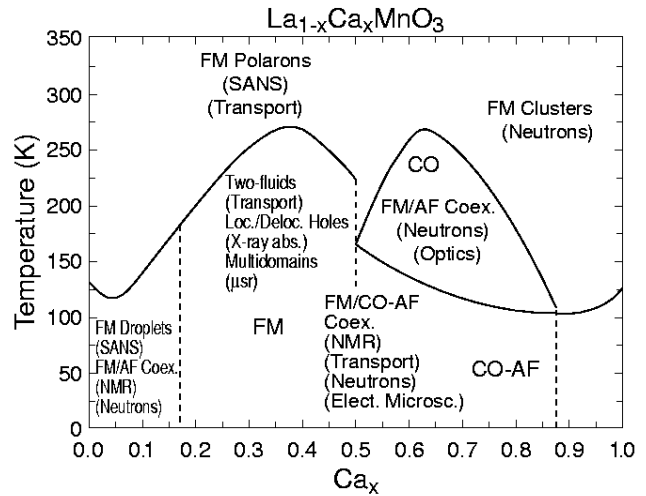,width=9.9cm,angle=-90}}
\vspace{0.1cm}
\caption{ 
A schematic version of the phase diagram for
$\rm La_{1-x} Ca_x Mn O_3$ (see Ref.~[3]).
Some of the experiments listed were performed for 
$\rm Bi_{1-x} Ca_x Mn O_3$, the bismuth analog. The
abbreviations Coex., Loc., Deloc., abs., $\rm \mu SR$ and EM, stand for
coexistence, localized, delocalized, absorption, muon spin relaxation,
and electron microscopy, respectively.
}
\end{figure}

For the large hole-density regime,
neutron scattering 
by Bao et al.~[50] at $\rm x$$\sim$$0.8$ 
using $\rm Bi_{1-x} Ca_x Mn O_3$, which
behaves similarly to $\rm La_{1-x} Ca_x Mn O_3$,
found FM-AF features coexisting between 150 and 200K.
Optical measurements at
$\rm x$$>$$0.5$ by Liu et al.~[51]  reported
similar features. Cheong
and Hwang in [3] found  a finite magnetization 
at low-T and $\rm x$$~\geq$$0.83$ 
 in the $\rm (La,Ca)$-compound [see also~(52)]. The system remains
insulating, and the results could be compatible
with spin-canted or mixed-phase states. 
More work at small electronic density
is needed to clarify if the phase segregation 
predicted by the theoretical calculations indeed
appears in experiments.

\vskip 0.3cm

{\bf VI.3 Manganese Oxides with a Charge-Ordered State
Near x=0.5:}
Results involving the
 ``charge-ordered'' AF-state
at $\rm x$$\sim$$0.5$ require special
discussion. Here, the extraordinarily large CMR
effect involves the abrupt
destabilization of the CO-state by a magnetic field~[5].
Evidence for PS tendencies is rapidly
accumulating in this region of the phase diagram of
narrow band manganese oxides.
Several experiments for $\rm La_{1-x}Ca_x Mn O_3$ 
have already reported coexisting
 metallic FM and insulating CO-AF clusters near
$\rm x$$=$$0.5$ (Fig.5)~[53].
Another compound at the FM-CO boundary 
at low-T is $\rm Pr_{0.7} Ca_{0.3} Mn O_3$. Here 
x-ray synchrotron and neutron-powder diffraction results~[54,55] 
were attributed to the presence of
 FM clusters in the CO-phase. Exposure to x-rays 
produces a nonuniform PS
phenomenon characteristic of two competing states, 
with an increasing size of the FM droplets and no evidence of 
spin-canting. 
This phenomenon is expected to appear at
other hole densities as well.
Related manganese oxides exhibit similar features.
For example,
the absorption spectra of
thin films of $\rm Sm_{0.6} Sr_{0.4} Mn O_3$ have 
been attributed~[32]
to the formation of large clusters
 of the CO-state above  its ordering temperature.
In $\rm (La_{0.5} Nd_{0.5})_{2/3} Ca_{1/3} Mn O_3$,
insulating 
CO and metallic FM regions coexist~[56].

However, consider $\rho$ at 300~K for
$\rm La_{1-x} Ca_x Mn O_3$~[3].
A smooth
connection between the undoped, lightly-doped, and heavily-doped 
compounds seems to exist even in this narrow-band material. 
No obvious precursors for
$\rm x$$\geq$$0.5$ of the low-T CO-state have been reported.
The same occurs for
$\rm Nd_{0.5} Sr_{0.5} Mn O_3$, which is metallic above $\rm T_C$, 
becomes FM upon cooling, and reaches
the CO-state through a first order
transition upon further decreasing T~[57]. 
These results establish a possible qualitative
difference between the MR effect at small x and $\rm x$$\sim$$0.5$.
In the former, charge inhomogeneities appear above and below the
critical temperatures, and the mutual influence 
of neighboring phases (notably FM and PS) is important.
However, at $\rm x \sim 0.5$, 
the CO- and FM-states do not seem to have much influence on each other.
It could even be that 
the low-T 
microscopic PS tendencies at $\rm x$$\sim$$0.5$
may be caused by metastabilities rather than
a stable CI-state. 
However, since the 
competing states at $\rm x \sim 0.5$ have similar $\rm \langle n \rangle$s
charge inhomogeneities involving large clusters are possible
because Coulomb interactions will not prevent it (see above).
Further experimental work is needed to clarify this situation.
The theoretical study of competing FM and CO-AF states 
of narrow band manganese oxides in a single formalism 
also represents a 
challenge for computational studies. Preliminary results are
promising since a CE-type CO state has been recently stabilized in
computer simulations carried out at large electron-phonon coupling~[58].

\vskip 0.3cm
{\bf VI.4 Layered Manganites:}
Tendencies toward PS in layered manganites have also been observed.
Neutron scattering results~[59] 
for the bilayered $\rm La_{1.2} Sr_{1.8} Mn_2 O_7$ 
revealed a weak peak at the 
AF-momentum of the parent compound
that coexists with the dominant FM signal~[60].
Recently, PS between A-type
metallic and CE-type insulating CO-states was reported
for $\rm La_1 Sr_2 Mn_2 O_7$~[61].
In addition, studies of the one-layer material
$\rm Sr_{2-x} La_x Mn O_4$~[62] observed direct evidence 
for macroscopic
PS at small electronic density~[63].

\vskip 0.5cm

{\bf VII. The PS scenario 
compared with other theories for manganites.}

\vskip 0.2cm

 The PS scenario is qualitatively
 different from other theories proposed to explain the CMR effects in 
manganese oxides. It improves on the simpler versions of the DE ideas~[6] by
 identifying charge inhomogeneities as
 the main effect competing with ferromagnetism, and by
 assigning the insulating properties above $\rm T_C$,
 fundamental for the low hole-density CMR effect, to the influence
 of those competing phases.
In particular, 
the compressibility increase
above $\rm T_C$ caused by PS
leads to dynamical cluster formation.

 The ideas presented here also differ qualitatively from those
by Millis et al.~[23], although there are common
aspects. In the PS scenario, charge inhomogeneities over several lattice
spacings, not contained in local mean-field approximations~[23], 
 are believed to be relevant
for the description of the insulating state above $\rm T_C$. 
In addition, the orbital ordering 
plays a key role in the results presented in Fig.1.
Although the importance of the
JT-coupling introduced
by Millis et al.~[23] is shared in both approaches, in the PS
scenario a state formed by independent local polarons is a special case of a more general
situation where clusters of various sizes and charges are possible.
These fluctuations increase as $\rm T_C$ decreases explaining
the optimization of the MR effect at the boundary of the FM-phase.
Note that the regime of small x is crucial to distinguish between the PS 
scenario and other polaronic theories
based on more extended polarons
and percolative processes~[31].

Other theories are based on the electronic localization effect
using the off-diagonal disorder intrinsic to the 
DE model~[64], and nonmagnetic diagonal disorder caused by the 
chemical substitutions.
These effects lead to a large MR under some approximations~[65]. 
However, 
the calculations are difficult because they
involve both strong couplings and disorder, and the
prominent cluster
and polaron formation found in experiments has not been addressed in
this framework. 
A better starting point for manganites may need
a formalism that accounts for the tendency to develop
 charge inhomogeneities
before including disorder.

\vskip 0.5cm
{\bf VIII. Conclusions}
\vskip 0.2cm

A variety of recent calculations have found PS tendencies
in models for manganites, usually involving FM and AF phases. 
These tendencies, which should lead
to a stable but microscopically 
inhomogeneous state upon the inclusion of Coulomb
interactions,  compete strongly with
ferromagnetism in the phase diagrams and are expected to increase
substantially the resistivity. 
Particularly, when two-orbital models are studied,
the results are in good
agreement with a large list of experimental observations 
reviewed here. 
Tendencies toward charge inhomogeneous states exist in
real manganese oxides all around 
the FM-phase in the temperature-density phase diagram.
The computer simulations have shown that the region with PS
 tendencies substantially influences the stable FM-phase by
increasing its compressibility, an aspect that
can be tested
experimentally. 
This provides a rationalization for the  
experimental observation of a large
MR effect at the boundaries of the FM-phase.
The presence of short-range charge
correlations is certainly a crucial feature of the PS scenario.

However, considerable work still remains to be done. The inclusion of extended
Coulomb interactions and the stabilization of the
x=0.5 charge-ordered
state are the next challenges for computational studies. Analytical
techniques beyond the local mean-field approximations are needed to
capture the essence of the charge inhomogeneous state.
Macroscopic phenomenological approaches should be used to obtain
predictions for transport properties and the shapes of the clusters
that arise from the competition of the DE attraction and Coulomb
repulsion. 
These resulting clusters are surely not static but fluctuating,
specially above $\rm T_C$.
In related problems of nuclear physics at high density,
several geometries have been found including spherical drops, rodlike
structures (stripes~[28] or ``spaghetti'') and platelike ones
(``lasagna'')~[66].
Similar rich phenomena may occur in manganites.
On the
experimental front it is crucial to establish if the various
regimes with charge inhomogeneities (Figs.4B and 5) are 
related, as predicted by the theoretical calculations.
For example, work should be carried out to link the small x
regime of  $\rm
(La,Ca)$-manganites where FM-droplets appear,
with the polarons reported at the x=1/3 density, and
beyond into the highly hole-doped regime. In addition, phase segregation
tendencies should also be studied close to the fully doped limit $\rm \langle n
\rangle \ll 1$ of manganese oxides, as well as
in related compounds such as doped
AF semiconductors~[67].

\vskip 0.5cm
{\bf Acknowledgments}
\vskip 0.2cm

Part of the ideas discussed 
here were developed in collaborations with 
N. Furukawa, J. Hu, and A. Malvezzi~[9].
The authors are specially thankful to
W. Bao, S. J. Billinge, C. H. Booth, S. L. Cooper, T. Egami, A. Fujimori,
 N. Furukawa, J. Goodenough, M. Hennion,
M. Jaime, J. Lynn, S. Maekawa,
Y. Moritomo, J. Neumeier, T. W. Noh, 
G. Papavassiliou, P. G. Radaelli, A. Ramirez, P. Schiffer,
Y. Tokura, H. Yoshizawa for their important comments and criticism.
A. M. and E. D. are supported in part by grant NSF-DMR-9814350.

\medskip


\medskip

\end{document}